\documentclass[aps,superscriptaddress,showpacs,preprintnumbers,nofootinbib,amsmath,amssymb,]{revtex4}
\usepackage{braket}
\usepackage{bm}
\usepackage{dcolumn}
\usepackage{hyperref}
\usepackage[dvipdfmx,final]{graphicx}
\usepackage{ulem}
\input{colordvi.tex}
\usepackage{subfigure}
\usepackage{verbatim}
\usepackage{makeidx}
\usepackage{amssymb}
\usepackage{float}
\usepackage{url}
\usepackage{color}
\newcommand{\be}{\begin{eqnarray}}
\newcommand{\ee}{\end{eqnarray}}

\newcommand{\Mpc}{\,{\rm Mpc}}
\newcommand{\GeV}{\,{\rm GeV}}
\newcommand{\eV}{\,{\rm eV}}
\newcommand{\mK}{\,{\rm mK}}

\begin{document}
\title{Impacts of new small-scale N-body simulations on dark matter
  annihilations constrained from cosmological 21cm line observations}

\author{Nagisa Hiroshima,}
\email{hirosima@sci.u-toyama.ac.jp}
\affiliation{Department of Physics, University of Toyama, 3190 Gofuku, Toyama 930-8555, Japan}
\affiliation{RIKEN Interdisciplinary Theoretical and Mathematical Sciences (iTHEMS), Wako, Saitama 351-0198, Japan}

\author{Kazunori Kohri,}
\email{kohri@post.kek.jp}
\affiliation{Theory Center, IPNS, KEK, 1-1 Oho, Tsukuba, Ibaraki 305-0801, Japan}
\affiliation{The Graduate University for Advanced Studies (SOKENDAI), 1-1 Oho, Tsukuba, Ibaraki 305-0801, Japan}
\affiliation{Kavli IPMU (WPI), UTIAS, The University of Tokyo, Kashiwa, Chiba 277-8583, Japan}

\author{Toyokazu Sekiguchi}
\email{tsekiguc@post.kek.jp}
\affiliation{Theory Center, IPNS, KEK, 1-1 Oho, Tsukuba, Ibaraki 305-0801, Japan}

\author{Ryuichi Takahashi}
\email{takahasi@hirosaki-u.ac.jp}
\affiliation{Faculty of Science and Technology, Hirosaki University, 3 Bunkyo-cho, Hirosaki, Aomori 036-8561, Japan}

\date{\today}

\begin{abstract} 
  We revisit constraints on annihilating dark matter based on the
  cosmological global 21cm signature observed by EDGES.  For this
  purpose, we used the numerical data of the latest N-body simulation
  for the first time performed by state-of-the-art standard in order to
  estimate the boost factor at high redshifts ($z$ = 10 -- 100), which
  enhances the annihilation of dark matter in course of structure
  formations.  By taking into account to what fraction injected energy
  from dark matter annihilation contributes to ionization, excitation,
  and heating of intergalactic medium during dark ages, we estimated
  how large the global 21cm absorption can be.  In the thermal
  freezeout scenario, we find that the dark matter masses
  $m_{\rm DM} < 15$~GeV and $m_{\rm DM} < 3$~GeV have been excluded at
  95$\%$ C.L. for the modes into $b\bar{b}$ and $e^+ e^-$,
  respectively, which are obtained independently of any uncertainties
  in local astrophysics such as observationally-fitted density
  profiles of dark matter halos.
\end{abstract}

\preprint{
KEK-TH-2314\quad
KEK-Cosmo-0275\quad
UT-HET-133\quad
RIKEN-iTHEMS-Report-21}
\maketitle

\section{Introduction} 
\label{sec:introduction}

In scenarios of dark matter (DM) for the weakly-interacting massive
particle (WIMP)~\cite{Jungman:1995df}, a DM particle with mass of the
weak scale should have a thermally-averaged annihilation cross-section
of the order of
$\langle \sigma v \rangle = 2 \times 10^{-26} {\rm cm}^3/{\rm sec}$ 
\cite{Steigman:2012nb,Saikawa:2020swg}.
This value is called the ``canonical annihilation cross-section'' and
is obtained to agree with the observed abundance of DM 
which was produced from the thermal bath in the early Universe by the
thermal freezeout mechanism (e.g, see Ref.~\cite{Saikawa:2020swg} and
references therein). Nowadays, a lot of new experimental projects have
been proposed to observe signatures of annihilating DM, 
which heightens the momentum toward cosmologically verifying the existence
of the WIMP DM.

On the other hand, it is still an open question how density
fluctuations of DM 
can evolve nonlinearly at high redshifts
$z = {\cal O}(10) - {\cal O}(10^2)$ and form structures at small
scales down to $k \sim 10^6 - 10^7 {\rm Mpc}^{-1}$ because 
nonlinear evolution of the density fluctuations becomes important.
Actually
two of the authors (KK and RT) of this paper have performed the
detailed N-body simulation of cold DM 
(CDM) in a separate paper~\cite{Takahashi:2021pse}. As a result, they are suggesting that the
halo formation is really sizable even at such a high redshift epoch.
The results of Ref.~\cite{Takahashi:2021pse} enables us to calculate the boost factor as a function of redshift with a sufficient level of precisions. 
It is a striking point that
this result means that DM 
could have inevitably annihilated
at the high redshifts, and the energy injection by the annihilating DM 
is expected to be the order of
${\cal O}(10^{-21}) \mathrm{eV} / \mathrm{sec} / \mathrm{cm}^{3}$ at
$z\sim 20$, which affects the absorption feature of the global 21cm
line-spectrum (e.g., see
Refs.~\cite{Poulin:2016anj,DAmico:2018sxd,Mena:2019nhm,Liu:2020wqz,Bolliet:2020ofj}
and references therein).

So far, the Experiment to Detect the Global Epoch of Reionization Signature (EDGES) 
collaboration reported the observational data for
the absorption feature of the cosmological global 21cm
line-spectrum~~\cite{Bowman:2018yin} at around $z\sim 17$. In a
pioneering work~\cite{DAmico:2018sxd}, by using this data the authors
obtained upper bounds on the annihilation cross-section of DM 
in order not to reduce the trough of the absorption feature due to the 
extra heatings by the annihilations. The boost factors adopted
in~\cite{DAmico:2018sxd} were taken from the values from the following
two papers with assuming somehow aggressive levels of model-dependent
approximations: i) Ref~\cite{Ripamonti:2006gq} in which the
Press-Schechter mass function formalism~\cite{Press:1973iz} with the Navarro-Frenk-White
(NFW) halo profile~\cite{Navarro:1996gj} was assumed, and ii) Ref.~\cite{Liu:2016cnk} in
which a halo model with the Einasto profile~\cite{Einasto:1965czb} was assumed. 
\footnote{
About other
related works to constrain DM 
from the data of EDGES, see also the 
following papers and references therein, Refs.\cite{Yang:2018gjd,Jia:2019yhr,Liu:2020wqz} for
annihilations, 
Refs.\cite{Clark:2018ghm,Mitridate:2018iag,Liu:2020wqz,Bolliet:2020ofj}
for decays, and Refs.~\cite{Tashiro:2014tsa,Xu:2018efh,Munoz:2018pzp,Slatyer:2018aqg} for DM-baryon interactions. And see also Refs.~\cite{Pospelov:2018kdh,Moroi:2018vci,Yoshiura:2018zts} for productions of additional photons to fit the EDGES data. 
}

\footnote{
  And also see Refs.~\cite{Nakama:2017qac} for constraints on the
  curvature perturbation at small scales from gamma-ray and neutrino
  observations produced in ultracompact minihalos of annihilating dark
  matter at present, which were calculated analytically by the
  Press-Schechter formalism while keeping the annihilation cross
  section to be the canonical one.  }

On contrary to those previous works, we compute the boost factor as a
function of $z$ by adopting the raw numerical data of the detailed
N-body simulations at the small scales reported
in~\cite{Takahashi:2021pse}. By using this boost factor, we update the upper
bounds on annihilating DM 
with the modes into $W^+ W^-$, $b\bar{b}$, $e^+ e^-$ and $\gamma \gamma$ as conservatively
as possible.

This paper is organized as follows. 
In Section \ref{sec:igm} we review how energy injection in general affects the evolution of the intergalactic medium (IGM). 
In Section \ref{sec:injection} we review how high energetic injection from DM 
annihilation is deposited into IGM. In Section \ref{sec:nbody} we describe 
how we estimate the annihilation boost factor based on our dedicated N-body simulation. 
In Section \ref{sec:results} we present our results. We conclude in the final section.

\section{Evolution equations of IGM in the presence of energy injection} 
\label{sec:igm}
In this section, we review how energy injection in general takes part in evolution of IGM.
For illustrative purpose, we here follow the simple description of hydrogen in IGM based on the effective three-level
atom model\cite{Peebles:1968ja,Zeldovich:1969en,Seager:1999km}. In numerical calculation we present in Section \ref{sec:results}, we adopt the recombination code {\tt HyRec}\footnote{\url{https://pages.jh.edu/~yalihai1/hyrec/hyrec.html}}, which is based on the state-of-the-art effective multi-level atom model (See~\cite{AliHaimoud:2010dx,Chluba:2010ca} for details). We in this paper focus only on hydrogen ionization and recombination assuming helium is neutral, which should be a good approximation as long as we are interested in the Dark Ages~\cite{Liu:2016cnk} (See also \cite{Liu:2019bbm}). 

The evolution of ionization fraction, $x_e$, is then described by the following equation:
\be
\frac{dx_e}{dt}&=&-C\left[\alpha_{\rm H}(T_m)x_e^2 n_H-\beta_{\rm H}(T_\gamma)(1-x_e)
e^{-E_\alpha/T_\gamma}\right]\notag\\
&&\quad+\frac{dE_{\rm inj}}{dVdt}\frac1{n_{\rm H}}\left[
\frac{f_{\rm ion}(t)}{E_0}+\frac{(1-C)f_{\rm exc}(t)}{E_\alpha}
\right], \label{eq:dotxe}
\ee
where $T_m$ and $T_\gamma$ are respectively the temperatures of gas and photon, $n_{\rm H}$ is the number density of hydrogen, $E_0\simeq13.6\eV$ is the ionization energy of hydrogen, $E_\alpha=3E_0/4$
is the energy of Ly-$\alpha$ photon, $\alpha_{\rm H}$ is the case-B recombination coefficient and $\beta_{\rm H}$ is the corresponding ionization rate. 
The Peeble's $C$-factor, which represents the probability that a hydrogen atom initially in the $n=2$ 
shell reaches the ground state without being photoionized, 
is given by
\be
C=\frac{\Lambda n_{\rm H}(1-x_e)+\frac1{2\pi^2}E_\alpha^3 H(t)}
{\Lambda n_{\rm H}(1-x_e)+\frac1{2\pi^2}E_\alpha^3 H(t)+\beta_Hn_H(1-x_e)},
\ee
where $\Lambda\simeq8.23$s$^{-1}$ is the two-photon decay rate of the hydrogen 2$s$-state,
and $H(t)$ is the Hubble expansion rate. The last term in \eqref{eq:dotxe} represents the effects of
energy injection, which we will describe shortly after.

The evolution of the gas temperature $T_m$ is described by the following equation:
\be
\frac{dT_m}{dt} &=&
-2H(t)T_m+\Gamma_C (T_{\gamma}-T_m)
+\frac{dE_{\rm inj}}{dVdt}\frac1{n_{\rm H}}
\frac{2f_{\rm heat}(z)}{3(1+x_e+f_{\rm He})}, \label{eq:dotTm}
\ee
where  $\Gamma_C$ is the coupling rate of $T_m$ to $T_\gamma$, which is predominated by the Compton scattering, 
\be
\Gamma_C=\frac{8\sigma_Ta_rT_\gamma^4}{3m_e} \frac{x_e}{1+f_{\rm He}+x_e},
\ee
where $\sigma_T$ is the Thomson scattering cross section, $a_r$ is the radiation constant, $m_e$ is the electron mass, and $f_{\rm He}$ is the number ratio of helium to hydrogen. 

The last term in each of Eqs. \eqref{eq:dotxe} and \eqref{eq:dotTm}, which is proportional to the
energy injection rate per unit volume per time, $dE_{\rm inj}/(dVdt)$, 
represents the effect
of energy injection.  As defined in \cite{Slatyer:2015jla,Slatyer:2015kla} the coefficients 
$f_{\rm ion}(t)$, $f_{\rm exc}(t)$, and $f_{\rm heat}(t)$ (collectively denoted by $\{f_c(t)\}$ hereafter) 
are the fractions of injected energy deposited into the hydrogen ionization, the hydrogen excitation 
and the heating of gas, respectively, which will be discussed in the next section.

\section{Energy injection and deposition into IGM from DM 
annihilation}
\label{sec:injection}
In the case of DM 
annihilation, the energy injection rate is given as\footnote{
Here DM 
is assumed to be self-conjugate.
}
\be
\frac{dE_{\rm inj}}{dVdt} = 
\bar\rho_{\rm DM}^2 B(z) \frac{\langle \sigma v\rangle}{m_{\rm DM}},
\ee
where $\bar \rho_{\rm DM}$ is the mean energy density of DM. 
$B(z)=\langle \rho_{\rm DM}^2 \rangle/{\bar \rho_{\rm DM}}^2$ 
is the boost factor due to the inhomogeneity 
of DM 
distribution, which will be discussed
in Section \ref{sec:nbody}, $\langle \sigma v\rangle$ is the annihilation cross-section 
averaged over the phase space distribution
and $m_{\rm DM}$ is the DM 
mass.

 The deposition fractions $\{f_c(t)\}$ depend on particle constituents and their energy spectra from 
DM 
annihilation as well as their interaction with IGM. We compute  $\{f_c(t)\}$ stepwise as follows:
\begin{enumerate}
\item  Once the primary annihilation processes (e.g. ${\rm DM}+\overline{\rm DM}\to{\rm SM}+\overline{\rm SM}$) are specified, the standard model (SM) particle   constituents and their energy spectra of the final state can be computed based on Monte Carlo event generators (e.g. {\tt PYTHIA}\footnote{\url{http://home.thep.lu.se/~torbjorn/Pythia.html}} and {\tt HERWIG}\footnote{\url{https://herwig.hepforge.org}}) which can simulate cascades of primary annihilation products into stable SM particles. In this paper, 
for  $m_{\rm DM}$ above $5\GeV$ we adopt  {\tt PYTHIA} to compute the energy spectra (For details, we refer to
\cite{Sjostrand:2007gs,Bahr:2008pv,Sjostrand:2014zea} and reference therein).

On the other hand, 
for $m_{\rm DM}$ below $5\GeV$, where we in this paper restrict ourselves to primary annihilation channels into $e^+e^-$ and $\gamma\gamma$,
we omit final state radiations and adopt monochromatic energy spectra from the primary annihilation processes.
Since the fraction of energy carried by final state radiations is small and primary annihilation products efficiently deposit their energy into IGM at low energy, 
our treatment should be a good approximation.

\item 
Energetic electrons, positrons and photons ejected from DM 
annihilation subsequently interact with IGM. 
How those energetic particles lose their energy through interaction with IGM and affect ionization and heating of IGM have been 
studied by many authors, e.g., \cite{Shull:1985,Chen:2003gz,Padmanabhan:2005es,Ripamonti:2006gq,Kanzaki:2008qb,
Slatyer:2009yq,Kanzaki:2009hf,Evoli:2012zz}. Energetic electrons/positrons lose their energy on timescales shorter than the Hubble time. Meanwhile, the timescale of energetic photons above $\simeq10^3\eV$ and below $\simeq10^{11}\eV$ can be longer than the Hubble time, which requires detailed computation of energy deposition over cosmological timescales. In this paper, we adopt the results of \cite{Slatyer:2015kla}\footnote{\url{https://faun.rc.fas.harvard.edu/epsilon/}}, which treats the effects of energy injection at linear level. For full treatments including feedback of modification of IGM evolution in computation of $\{f_c(t)\}$ we refer to \cite{Liu:2019bbm}. \\
\end{enumerate}

Analytically, we can estimate the energy injection rate to be
\begin{equation}
\begin{aligned} \frac{dE_{\rm inj}}{dVdt} \sim & 10^{-21} \mathrm{eV} / \mathrm{sec} / \mathrm{cm}^{3} \\ &
\times \left( \frac{B(z)}{10^2} \right) \left(\frac{1+z}{18}\right)^{6}\left(\frac{\langle\sigma v\rangle}{2 \times 10^{-26} \mathrm{~cm}^{3} / \mathrm{sec}}\right)\left(\frac{\Omega_{\mathrm{DM}} h^{2}}{0.12}\right)^{2}\left(\frac{m_{\mathrm{DM}}}{10^{2} \mathrm{GeV}}\right)^{-1}. \end{aligned}
\end{equation}
This order-of-magnitude energy injection rate can affect the absorption feature of the global 21cm line
spectrum~\cite{Poulin:2016anj,DAmico:2018sxd,Mena:2019nhm,Liu:2020wqz,Bolliet:2020ofj}.

\section{N-body simulation and the annihilation boost factor}
\label{sec:nbody}

DM 
annihilation is enhanced by inhomogeneity in DM 
distribution, which can be encapsulated in the boost factor $B(z)$. Even at redshifts as high as $z\gtrsim 15$ which we are focusing on in this paper, DM 
fluctuations at small scales have grown to be nonlinear. Therefore, to estimate $B(z)$, one needs to trace the nonlinear evolution of DM 
fluctuations. N-body simulations have been a powerful tool for this purpose. 

Denoting $\delta(\mbox{\boldmath{$x$}};z)$ as the DM 
density contrast at a spatial position $\mbox{\boldmath{$x$}}$ at redshift $z$, the boost factor is defined as $B(z)=1+\langle \delta^2 (\mbox{\boldmath{$x$}};z) \rangle$.  
It can be recast using the Fourier transform as \cite{Serpico2012,Sefusatti2014}
\be
B(z)=1+\int^\infty_0 \frac{dk}k\Delta^2(k;z),
\ee
with $\Delta^2(k;z) \equiv k^3 P(k;z)/(2 \pi^2)$ where $P(k;z)$ is the power spectrum of DM 
density fluctuations. 
The dimensionless power spectrum $\Delta^2(k;z)$ is almost flat at $k \gtrsim 10 \, \Mpc^{-1}$, where horizon crossing takes place during the radiation dominated era. 
Therefore, the expression indicates that $B(z)$ is contributed from a wide range of scales. This necessitates that N-body simulations should be performed with a variety of box-sizes, enabling DM 
fluctuations to be resolved at relevant scales. However, so far there have been few studies performing N-body simulations as such focusing on redshifts of our interest. In \cite{Takahashi:2021pse}, some of the authors of this paper have addressed this issue by performing dedicated N-body simulations.

As presented in \cite{Takahashi:2021pse}, a suite of cosmological N-body simulations with a variety of box-sizes (i.e., side lengths of cubic boxes) is performed.
The box-sizes range from $1 \, {\rm kpc}$ to $10 \, {\rm Mpc}$ to cover a wide range of scales, $k \simeq 1$--$10^7 \, \Mpc^{-1}$. 
The simulations are comprised of $2560^3$ collisionless particles.
The initial linear power spectrum is prepared using the transfer function \cite{Yamamoto1998} with the free streaming damping of DM 
particles at $k_{\rm fs}=10^6 \, \Mpc^{-1}$ \cite{Green2004}.
Initial conditions of the simulations are set at redshift $z=400$ based on the second-order Lagrangian perturbation theory \cite{Crocce2006,Nishimichi2009}. 
We employed the gravity solver {\tt GreeM} \cite{Ishiyama2009} to follow the nonlinear gravitational evolution.
To compute the boost factor, $\Delta^2(k)$ is constructed by connecting estimated DM 
power spectra at different wave number bands that depend on box-sizes of simulations. For more details of the simulations and analyses, we refer to \cite{Takahashi:2021pse}.

\begin{figure}
\centering
\includegraphics[width=8cm]{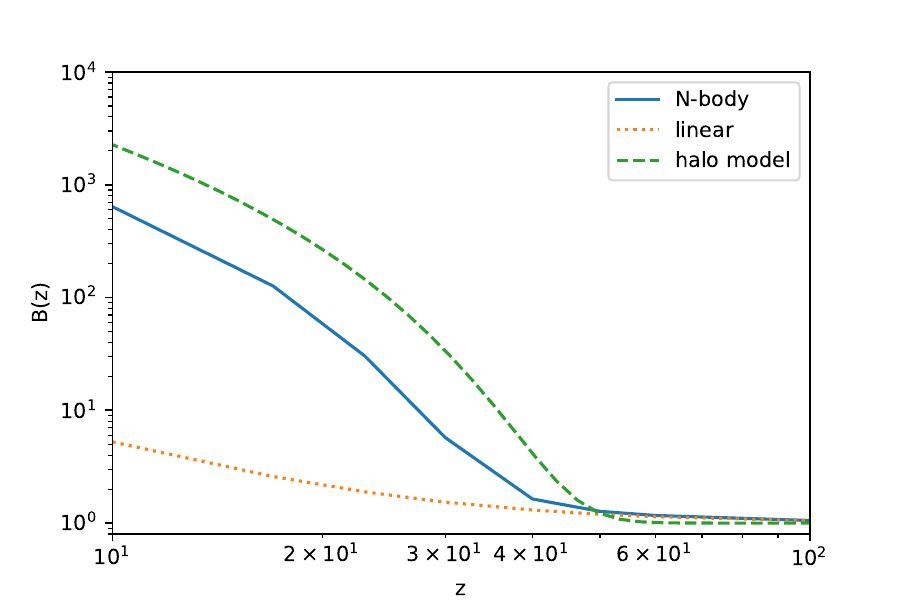}
\caption{\label{fig:B} Boost factors computed from linear perturbation
  calculations (orange dotted) and the N-body simulations (blue solid)
  done in Ref.~{\cite{Takahashi:2021pse}}. 
  For reference, we also depict $B(z)$ based on the halo model in \cite{Evoli:2014pva}, 
  which is referred to as the "Boost 1" model in \cite{DAmico:2018sxd} (green dashed).
}
\end{figure}

 Figure \ref{fig:B} shows $B(z)$ computed by using data of the simulations.
 As references, the figure also shows $B(z)$ computed based on linear perturbation theory 
as well as one based on the halo model in \cite{Evoli:2014pva}, 
which is referred to as the "Boost 1" model in \cite{DAmico:2018sxd}.

Here, in the linear theory, $B(z)-1$ simply evolves as $(1+z)^{-2}$.

Compared to other estimations, e.g. the "Boost 1" model adopted in \cite{DAmico:2018sxd} as a conservative choice,
our $B(z)$ is smaller than it at
$z\lesssim50$. This results in suppressed DM 
annihilation rate and hence may lead to much more conservative upper bounds on dark
matter annihilation cross-section.

\section{Results}
\label{sec:results}

\begin{figure}
\centering
\begin{tabular}{cc}
\includegraphics[width=7cm]{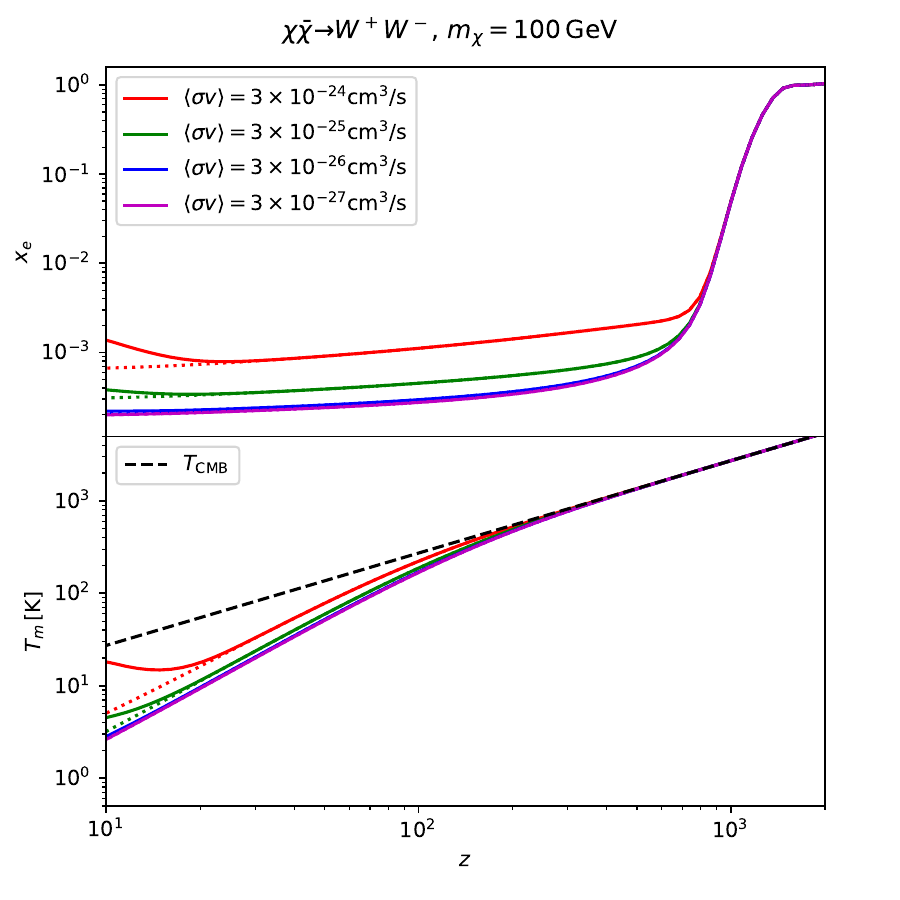} &
\includegraphics[width=7cm]{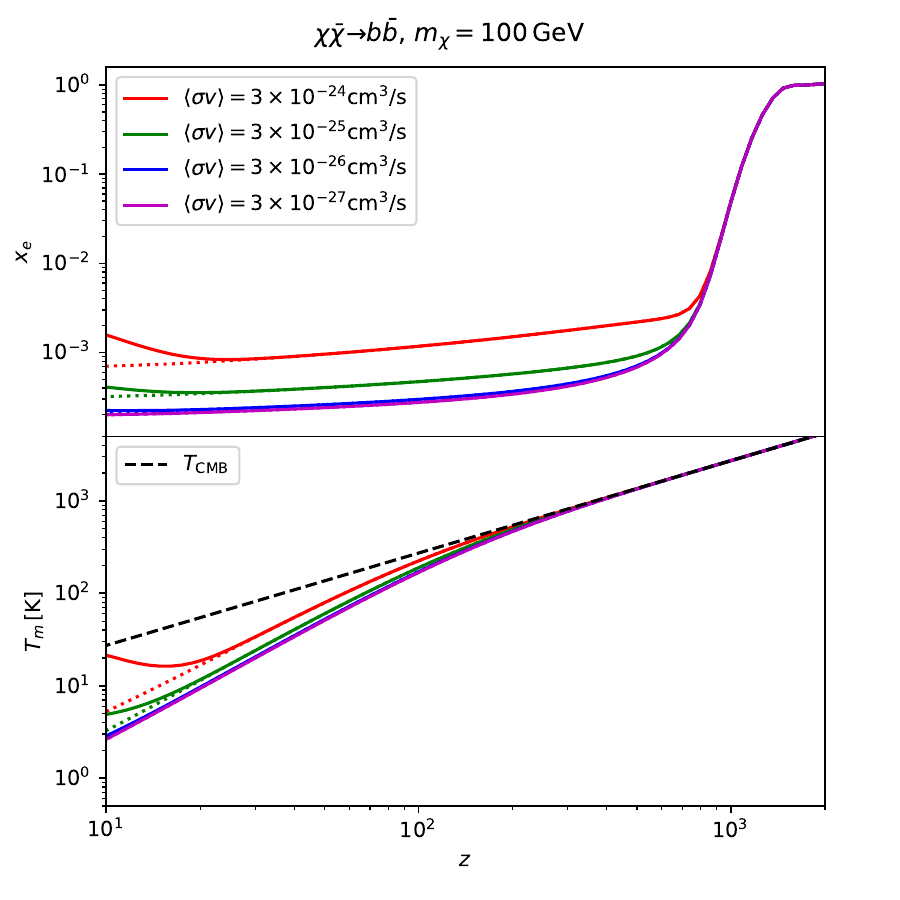} \\
\includegraphics[width=7cm]{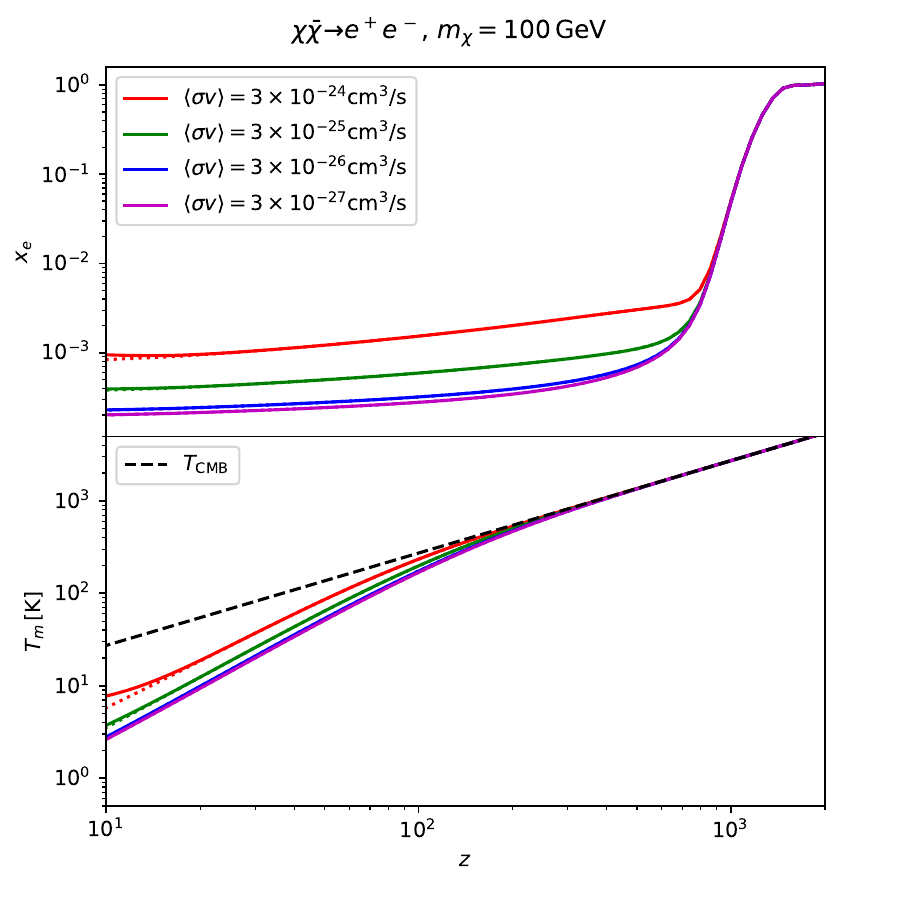} &
\includegraphics[width=7cm]{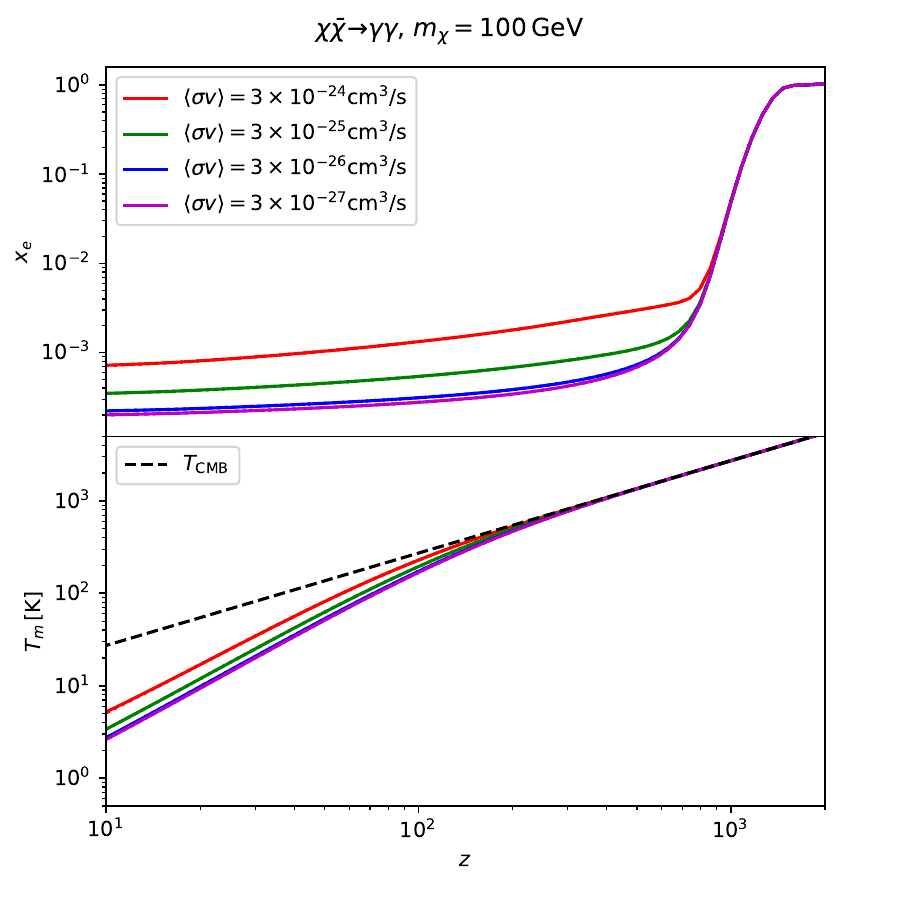} 
\end{tabular}
\caption{\label{fig:100GeV} Evolution of $x_e$ and $T_m$ in the
  presence of DM 
  annihilation.  From top left to bottom
  right, cases of annihilation channels into $W^+W^-$, $b\bar b$,
  $e^+e^-$, $\gamma\gamma$ are plotted. The DM 
  mass $m_{\rm DM}$ is assumed to be $100\GeV$. Annihilation cross-section
  $\langle \sigma v\rangle$\,[cm$^3$/s] is taken to be
  $3\times 10^{-24}$ (red), $3\times 10^{-25}$ (green),
  $3\times 10^{-26}$ (blue) and $3\times 10^{-27}$ (magenta). The
  boost factor $B(z)$ is computed based on the N-body simulation
  (solid) and the linear perturbation theory (dotted). For reference,
  $T_\gamma(z)$ (black dashed) is also plotted in each panel of
  $T_m(z)$. }
\end{figure}

In figure \ref{fig:100GeV}, we demonstrate the evolution of $x_e(z)$
and $T_m(z)$ in the presence of DM 
annihilation with DM 
mass $m_{\rm DM}=100\GeV$. We here assume that there is no
significant heating from astrophysical sources.  

The differences between the results of the linear theory and the N-body
simulation for the $W^+W^-$ and $b\bar{b}$ emissions are larger than
the ones for the line $e^+e^-$ and $\gamma\gamma$ emissions. That is
because more soft daughter electromagnetic particles such as $e^+e^-$
or $\gamma\gamma$ are produced through cascade decays of unstable
mesons and baryons in cases for the $W^+W^-$ and $b\bar{b}$ emissions,
compared with the cases for the line $e^+e^-$ and $\gamma\gamma$
emissions. Then, the energy-deposition is more efficient for $W^+W^-$
and $b\bar{b}$ due to the delayed deposition~\cite{Liu:2018uzy}.

$T_m$ is increased compared to cases where energy injection from dark
matter annihilation is absent, which should result in modified
evolution of the spin temperature, $T_s$, associated with the
hyperfine splitting of neutral hydrogen ground states. This allows us
to constrain DM 
annihilation cross-section from observations
of differential brightness temperature of redshifted 21\,cm line
emission \be T_{\rm 21cm}(z)=\frac{T_s(z)-T_\gamma(z)}{1+z} \tau_{\rm
  21cm}(z), \ee before reionization (See
e.g. \cite{Furlanetto:2006jb}).  The time-evolution of $T_s$ in general
depends on relative couplings of $T_s$ with $T_\gamma$, $T_m$ and the
color temperature $T_c$, which is the effective temperature associated
with background Lyman-$\alpha$ radiation. As IGM is always optically
thick for Lyman-$\alpha$ radiation during the cosmological epoch we
are interested in, it is reasonable to assume $T_c\approx T_m$. The
fact that EDGES has reported a global absorption
signal~\cite{Bowman:2018yin} \be T_{\rm 21cm}=-
500^{+200}_{-500}\,{\rm mK}\quad(\mbox{99\% CL}), \ee indicating
$T_m<T_\gamma$, then the upper bounds on the DM annihilation cross-section should exist. 
This is because DM 
annihilation in general suppresses the absorption
amplitude of global 21cm signals by ionizing and heating IGM. In
Figure~\ref{fig:100GeV}, the differences between the solid and dotted
lines look much smaller than the ones directly-expected from the face values of the difference between the two lines in
the boost factors from our N-body simulations and linear perturbation calculations in
Figure~\ref{fig:B}.  That is because the delayed deposition occurred as
was discussed in Ref.~\cite{Liu:2018uzy,Basu:2020qoe}.

\begin{figure}
\centering
\includegraphics[width=10cm]{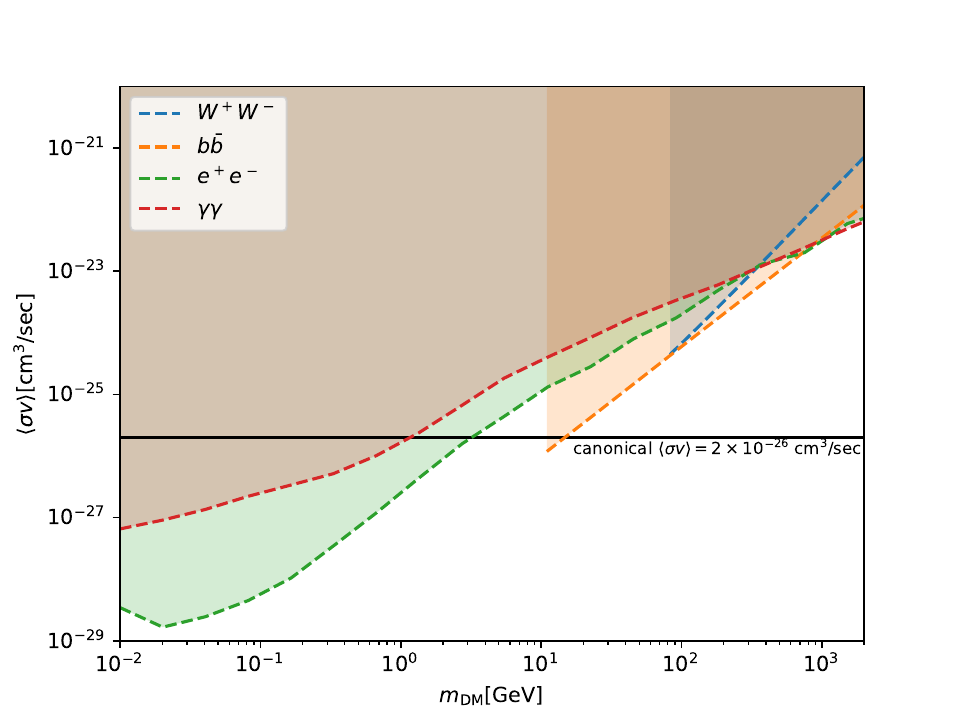}
\caption{\label{fig:constraints} Constraints on DM 
annihilation cross-section for the modes into $W^+ W^-$ (blue), 
$b\bar{b}$ (orange), $e^+ e^-$ (green) and $\gamma \gamma$ (orange).
We here put the upper bound by assuming 
 $T_{\rm 21cm}\le -75\mK$ corresponding to $2\sigma$ deviation from the baseline.
The canonical WIMP cross section $\langle \sigma v\rangle=2\times 10^{-26}\,{\rm cm}^3/{\rm sec}$ is also shown for reference (black solid line). 
}
\end{figure}

We in particular obtain conservative upper bounds on DM 
annihilation cross-section by maximizing the absorption amplitude in the absence of DM 
annihilation~\cite{DAmico:2018sxd}. This can be realized by assuming no heating of IGM other than DM 
annihilation. We also assume tight coupling of spin temperature to gas temperature via Lyman-$\alpha$ pumping (Wouthuysen-Field effect \cite{Wouthuysen:1952,Field:1959}), which can also maximize the absorption depth. Figure \ref{fig:constraints} shows upper bounds on DM 
annihilation cross-section based on this strategy.  Our baseline calculation in the absence of DM 
annihilation, gives $T_{\rm 21cm}\simeq - 230\,{\rm mK}$. We put upper bounds on the annihilation cross-section of the DM by
requiring $T_{\rm 21cm} \le -75\,{\rm mK}$, which correspond to the
2$\sigma$ bound given
uncertainties of EDGES.

From this figure, we find that the upper bounds on the annihilation
cross sections, which are conservatively-obtained in this study for the
$b\bar{b}$, $e^+e^-$ and $\gamma\gamma$ modes, are milder than the ones
in the results of~\cite{DAmico:2018sxd}. That is because the boost
factor we adopted is smaller than the ones
in~\cite{DAmico:2018sxd}. When we assume the canonical value of the
annihilation cross-section,
$\langle \sigma v \rangle = 2 \times 10^{-26} {\rm cm}^3/{\rm sec}$, we
can exclude the masses of DM 
for each mode to be $m_{\rm DM} < 15$~GeV ($b\bar{b}$), $m_{\rm DM} < 3$~GeV ($e^+ e^-$),
and $m_{\rm DM} < 1$~GeV ($\gamma \gamma$) at 95$\%$ C.L.

\section{Conclusion}\label{sec:conclusion}

In this paper we have revisited the possible constraints on
annihilation cross-sections of DM 
from the observations on
the cosmological global 21cm line-spectrum reported by EDGES. By
adopting the latest data of high-redshift dark-matter halo-formations
($z = {\cal O}(10) - {\cal O}(10^2)$) performed by the detailed N-body
simulations at the small scales, we have updated the boost factor of
the annihilating DM 
due to the clumpiness.

With this updated value of the boost factor, we obtained the more
conservative upper bounds on the annihilation cross-sections than the
ones reported in the previous work. In this study, we can exclude the
masses of DM 
for $m_{\rm DM} < 15$~GeV ($m_{\rm DM} < 3$~GeV) at 95$\%$ C.L. for
the mode into $b\bar{b}$ ($e^+ e^-$) by assuming the canonical value
of the annihilation cross-section,
$\langle \sigma v \rangle = 2 \times 10^{-26} {\rm cm}^3/{\rm sec}$. 
These bounds obtained from the global 21cm spectrum are cosmologically
robust along with the ones from CMB~\cite{Slatyer:2015jla} and
BBN~\cite{Kawasaki:2015yya} because they do not depend on local
astrophysical uncertainties.

In the future, we can improve sensitivities on the constraints on the
annihilation cross-section by adopting more precise observational data
which are expected to be reported by new projects such as
HERA~\cite{Beardsley:2014bea}, SKA~\cite{SKAspec},
Omniscope~\cite{Omniscope} or DAPPER~\cite{Burns:2021ndk}.

\acknowledgments We thank Sai Wang for useful discussions on early
stages of this work.  This research was supported by JSPS KAKENHI
Grant Numbers JP17H01131 (KK, TS and RT), JP15H02082 (TS), JP18H04339 (TS),
JP18K03640 (TS), JP19K23446 (NH) and MEXT KAKENHI Grant Numbers JP19H05114 (KK),
JP20H04750 (KK), JP20H05852 (NH), JP20H04723 (RT) and JP20H05855 (RT).
Numerical computations were carried out on Cray XC50 at Center for Computational 
Astrophysics, National Astronomical Observatory of Japan.

\appendix

\end{document}